\RequirePackage[l2tabu,orthodox]{nag}
\RequirePackage{fix-cm}
\documentclass[conference]{IEEEtran}
\IEEEoverridecommandlockouts

\usepackage{siunitx}
\usepackage{amsmath,amssymb,amsfonts}
\usepackage{algorithmic}
\usepackage{graphicx}
\usepackage{textcomp}
\usepackage{rotating}
\usepackage[english]{babel}
\usepackage{amsmath}
\usepackage[font=footnotesize,labelfont={bf,sc,footnotesize}]{caption}
\usepackage{dirtytalk}
\usepackage[utf8]{inputenc}
\usepackage[caption=false,font=footnotesize,labelfont=sf,textfont=sf]{subfig}
\usepackage[dvipsnames,svgnames,x11names,hyperref]{xcolor}

\usepackage[pdftex, breaklinks, colorlinks,citecolor=blue, bookmarks=false]{hyperref}
\hypersetup{ bookmarksnumbered=true,colorlinks=true, citecolor=red, bookmarksopen=true}

\PassOptionsToPackage{bookmarks=false}{hyperref}
\usepackage{enumitem}

\usepackage[ruled]{algorithm2e}
\newlength\mylen
\everymath\expandafter{\the\everymath\mathgroup0}
\everydisplay\expandafter{\the\everydisplay\mathgroup0}

\usepackage{booktabs, multicol, multirow}
\usepackage{colortbl}
\usepackage{rotating}
\usepackage{siunitx}
\usepackage{gensymb}
\usepackage{tabularx}
\usepackage{array}
\usepackage{makecell}
\usepackage{pbox}

\usepackage[activate={true,nocompatibility},final,tracking=true,kerning=true,spacing=true,factor=1100,stretch=10,shrink=10]{microtype}

\usepackage{cite}
\usepackage{amsmath,amssymb,amsfonts}
\usepackage{algorithmic}
\usepackage{graphicx}
\usepackage{textcomp}
\usepackage{xcolor}
\def\BibTeX{{\rm B\kern-.05em{\sc i\kern-.025em b}\kern-.08em
    T\kern-.1667em\lower.7ex\hbox{E}\kern-.125emX}}
        
\usepackage{fancyhdr}
     \usepackage[capitalise,noabbrev,nameinlink]{cleveref}
    \usepackage[symbol]{footmisc}

\begin{document}
	\setlength{\parskip}{0pt}
	\renewcommand*{\algorithmcfname}{Algorithm}
	\renewcommand*{\algorithmautorefname}{Algorithm}
	\renewcommand{\thefootnote}{\roman{footnote}}
	\renewcommand{\thefootnote}{\textcolor{Gray}{\arabic{footnote}}}
		
\def \datasetUrl {https://www.kaggle.com/qiriro/comfort}
\def \suppmaterial {\href{\datasetUrl }{\textcolor{Gray}{supplementary material}}}
\title{Affect-aware thermal comfort provision in intelligent buildings \vspace*{-4mm}}
\author{
	
	\IEEEauthorblockN{Kizito~Nkurikiyeyezu, Anna~Yokokubo, and Guillaume~Lopez}
	\vspace*{-4mm}
	\IEEEauthorblockA{\\ Wearable Information Lab \\
		Aoyama Gakuin University\\
		\{kizito, guillaume\}@wil-aoyama.jp, yokokubo@it.aoyama.ac.jp
	}
	\vspace*{-1cm}
}
\maketitle
\thispagestyle{fancy}
\begin{abstract}
	Predominant thermal comfort provision technologies are energy-hungry, and yet they perform crudely because they overlook the requisite precursors to thermal comfort. They also fail to exclusively cool or heat the parts of the body (e.g., the wrist,  the feet, and the head) that influence the most a person's thermal comfort satisfaction. Instead, they waste energy by heating or cooling the whole room. This research investigates the influence of neck-coolers on people's thermal comfort perception and proposes an effective method that delivers thermal comfort depending on people's heart rate variability (HRV).  Moreover, because thermal comfort is idiosyncratic and depends on unforeseeable circumstances, only person-specific thermal comfort models are adequate for this task. Unfortunately, using person-specific models would be costly and inflexible for deployment in, e.g., a smart building because a system that uses person-specific models would require collecting extensive training data from each person in the building.  As a compromise,  we devise a hybrid, cost-effective, yet satisfactory technique that derives a personalized person-specific-like model from samples collected from a large population. For example, it was possible to double the accuracy of a generic model (from 47.77\% to 96.11\%) using only 400 person-specific calibration samples. Finally, we propose a practical implementation of a real-time thermal comfort provision system that uses this strategy and highlighted its advantages and limitations.
\end{abstract}
\begin{IEEEkeywords}
	thermal comfort model, humanized computing, smart building, heart rate variability, energy conservation
\end{IEEEkeywords}
\section{Introduction}
Despite a century of research on thermal comfort, the technologies for its provision leave much to be desired \cite{Nicol2017,Brager2015}. By definition, thermal comfort is \say{the condition of mind that expresses satisfaction with the thermal environment and is assessed by subjective evaluation  \cite{ANSI/ASHRAE2013}}. Paradoxically, most thermal comfort provision technologies (e.g., air conditioning units) ignore this idiosyncratic nature of thermal comfort; instead, they provide neutral thermal conditions to all occupants of the buildings. Unfortunately, this strategy is inefficient and has many well-known flaws highlighted in, e.g., \cite{DeDear2013b,VanHoof2008}: First, a one-size-fits-all strategy cannot work well because of individual differences (e.g., age, gender, and physiological makeup) that influence how each person perceives thermal comfort \cite{Wang2018b}. Second, there is no rationale for providing \say{thermal neutral} conditions (i.e., conditions in which people feel neither warm nor cool\cite{DeDear2011a}). In reality, people prefer non-neutral conditions \cite{DeDear2011a, Humphreys2007, Brager1998}. Moreover, there is a mounting suspicion that thermal neutrality is pernicious because it may be a root cause of sick building syndrome (SBS) \cite{OleFanger2001, Kaczmarczyk2004}. Third, achieving thermal neutrality is costly and necessitates immoderate energy consumption \cite{Brager2015}. Forth, only a few parts of the body (e.g., head, wrists, and feet) are mostly responsible for thermal comfort. For example, in uniform environmental conditions, when it is cold, a person's feet and hands feel colder than other parts of the body. On the contrary, the cold environment does not affect the thermal sensation on the head, which usually feels warmer than the rest of the body and require a relatively lower temperature to achieve a satisfactory thermal comfort\cite{Arens2006}. However, air conditioning units do not exclusively direct the heat to these crucial parts of the body. Instead, they inefficiently cool or warm an entire room and regardless of the number of people available in the room. Lastly, current international thermal comfort standards are unambitious (they expect a mere 80\% satisfaction rate\cite{DeDear2011a}). Consequently, many luminaries in the field \cite{Nicol2017,DeDear2011a,Brager2015,VanHoof2008} argue for a paradigm shift in how thermal comfort is provided. 
\par 
In our previous research, we proposed a thermal comfort provision method that aimed to solve some of the above limitations. We argued that, since thermal comfort is a subjective psychological sensation and that thermoregulation leads to discernible physiological changes \cite{Morrison2011}, it would be more efficient to provide thermal comfort based on variations in a person's physiological signals. We revealed \cite{Nkurikiyeyezu2017a} that a change in a thermal environment led to detectable fluctuations in people's heart rate variability (HRV) and introduced an energy-efficient thermal comfort provision technique that uses people's physiological changes due to their surrounding thermal environments to provide a personalized thermal comfort \cite{Nkurikiyeyezu2018a}. We also developed a proof of concept machine learning-enabled apparatus that delivers, in real-time, a personalized thermal comfort \cite{Nkurikiyeyezu2018b, Lopez2018b, Lopez2018}. The system uses people's photoplethysmogram (PPG) signals to estimate their thermal comfort level. While we have not yet completed the development of the whole system, at its completion, we anticipate that the system shall deliver optimum thermal comfort based variation in people's physiological signals and that it shall minimize the energy consumption depending on, e.g., the number of people in the room, their comfort level, and depending on the outside weather. At the moment, this system performs crudely because it uses a generic thermal comfort prediction model and does not take into consideration the physiological differences between its users. In this paper, we aim to improve these limitations. 
\section{Methods} \label{sec:method}
\subsection{HRV datasets} \label{sec:dataset}
We used a thermal comforts dataset described in \cite{LOPEZ2016}. We collected the dataset by exposing eleven subjects to three experiments. In the first experiment, we placed the subjects in a very hot environment (32\textdegree{}C) and they wore adjustable (from 18\textdegree{}C up to 28\textdegree{}C) custom-made neck-coolers. In the second experiment, we put the subjects in the same very hot conditions, but in this case, they did not wear any neck-coolers. In the third experiment, we placed the subjects in a hot (29\textdegree{}C) environment and, like in the previous case, the subjects did not wear any neck-coolers. During the experiment, we recorded each subject's interbeat interval (IBI) using
myBeat heart rate sensor (Union Tool co.). We also recorded the subject's skin temperature on the chest, on the arm and the lower leg, on the neck and measured his/her sweating rate using SNT-200 sweat meters (Rousette Strategy, Inc.). The IBI signal was sampled at 1kHz. Other sensors were recorded at 1 sample per minute. The subject regularly self-evaluated their thermal comfort level, their thermal sensation, and their sweat level on a 10-scale  visual analog scale (VAS). All experiments lasted for at least 90 minutes.
\subsection{Feature extraction}  \label{sec:feature-computation}
\par 
We computed various heart rate variability (HRV) features
\footnote{\textcolor{Gray}{refer to Table I and to subsection I-A in the \suppmaterial}}

using the standards and algorithms proposed by the Task Force of the European Society of Cardiology and the North American Society of Pacing and Electrophysiology \cite{TaskForce1996}.  We first extracted an inter-beat interval (IBI) signal from the peaks of the electrocardiogram (ECG) signal of each subject. Then, we computed each HRV feature on a moving window as follows: We used a five-minute array of IBI to compute the first HRV index. Then, a new IBI sample is appended to the IBI array while the oldest IBI sample is removed from the beginning of the IBI array. The new resulting IBI array is used to compute the next HRV index. We repeated this process until the end of the entire IBI array. 
\subsection{Model training and evaluation} \label{model-validation}
We utilized the subject's self-assessment to develop regression models that estimate each subject's thermal comfort level. We also classified the individual's comfort based on the experiment conditions of their thermal environment. We used various machine learning algorithms (\autoref{tab:model-parameters}) to train and evaluate three thermal comfort prediction models: 
\begin{table}[htbp]
	\centering
	\caption{Thermal comfort models and their key hyperparameters}
	\begin{tabular}{@{}rrl@{}}
		\toprule
		Model &       & \multicolumn{1}{c}{Hyperparameters} \\
		\cmidrule{1-1}\cmidrule{3-3}    AdaBoost &       & estimator=DecisionTree(max\_depth=64), n\_estimators=50 \\
		Bagging &       & estimator=DecisionTree(max\_depth=64), n\_estimators=50 \\
		ExtraTrees &       & n\_estimators=100, max\_depth=64 \\
		RandomForest &       & n\_estimators=100, max\_depth=64 \\
		XGBoost &       & n\_estimators=50, max\_depth=64, subsample=0.8 \\
		\bottomrule
		
	\end{tabular}%
	\label{tab:model-parameters}%
\end{table}%

\begin{algorithm}[htbp]
	\DontPrintSemicolon 
	\SetAlgoLined
	\KwIn{machine learning algorithm $h_m$}
	\KwData{
		\begin{itemize}
			\item HRV samples $sample_{generic}$ of $n$ persons 
			\item Calibration HRV samples $sample_{calibration}$ that belong to $q$ unseen persons such that $q\ll n$
		\end{itemize}
	}
	
	\KwOut{trained calibrated model $h_m\prime$}
	\tcc{mix the calibration samples and the generic samples}
	$D^\prime \longleftarrow \emptyset$\;
	$D^\prime \longleftarrow shuffle(sample_{generic}\cup sample_{calibration})$\;
	\tcc{train the model  $h_m$ on dataset $D^\prime$}
	$h_m\prime \longleftarrow h_m(D^\prime)$\;
	\Return{$h_m\prime$}\;
	\caption{{\sc Model calibration} }
	\label{algo:model-calibration}
\end{algorithm}
\begin{itemize}
	\item \textit{generic model}\textemdash to assess how the model would perform in predicting the thermal comfort of new unseen subjects, (i.e., the subjects whose HRV samples were not part of the training set), we validated the performance of the generic model using a leave-one-subject-out (LOSO) approach. This method consists of using the data of one subject as a test set and the data of the remaining subjects as a training set. 
	\item  \textit{person-specific model}\textemdash unlike a generic model, a person-specific model is developed by training and testing the model exclusively on the data of the same person. We used a 10-folds cross-validation approach to evaluate the performances of each model.
	\item  \textit{hybrid model}\textemdash the generic model performs poorly because of individual differences in how people express thermal comfort. On the other hand, person-specific models would be costly to deploy in buildings because they would require acquiring and labeling the training data for each occupant of the building. We propose a hybrid method to mitigate this limitation. In a nutshell, the technique (\autoref{algo:model-calibration}) consists of adding a few person-specific samples (they are referred to as \say {calibration samples} in the remaining of this paper) collected from previously unseen people into a generic thermal comfort model trained on a large group of people. In this paper, we tested the method with $q=3$ on various algorithms (\autoref{tab:model-parameters}) \textemdash and their performances are similar. Nevertheless, all the relevant results in this paper are based on the predictions of Extremely Randomized Trees (ExtraTrees) models because they performed the best.
\end{itemize}

\section{Results}
\subsection{The effect of the neck-coolers on thermal comfort}
For most subjects, wearing the neck-coolers  improved how they felt about the thermal environments:
\begin{itemize}
	\begin{figure}[tbph]
		\centering
		\includegraphics[width=1.0\linewidth]{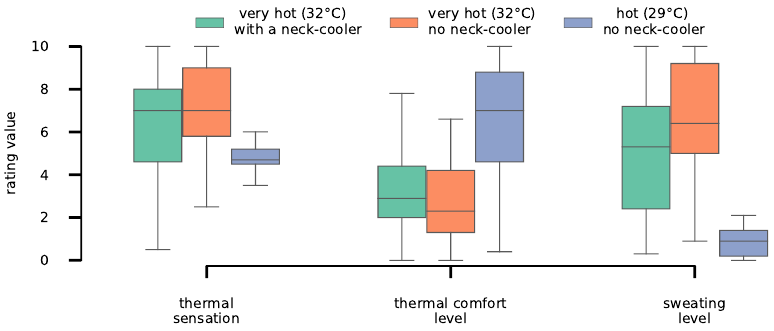}
		\caption{\textbf{Perceived thermal comfort and sweating rate} 
			\newline In general, the subjects(N=11 subjects) had a higher thermal comfort level and believed their sweating rate was lower when they wore the neck-coolers compared to when they did not wear any neck-coolers. However,  wearing the neck-coolers did not affect their thermal sensation significantly.}
		\label{fig:subjective-rating}
	\end{figure}
	\item As shown in \autoref{fig:subjective-rating}, the subjects expressed having a better thermal comfort level and a lower sweating rate when they wore the neck-coolers compared to when they did not, but the results of their thermal sensation is not conclusive (the median of the thermal sensation with or without the neck coolers is the same). Nevertheless, in all cases, wearing the neck-coolers was not sufficient to offset the impact of the 29\textdegree{}C temperature gap between the very hot and the hot environment.
	\item For all subjects, the skin temperatures and the neck temperatures decreased when they wore the neck-coolers $(\textit{P}<0.001)$. However, there is no conclusive evidence of the impact of the neck-coolers on the sweating rate because only the data of 8 subjects were statistically significant $(\textit{P}<0.05)$. Furthermore, similarly to the previous case, the influence of the neck-coolers was not strong enough to negate the temperature differences between a 32\textdegree{}C and a 29\textdegree{}C environment (\autoref{fig:personaltemperature}).
	\begin{figure}[htbp]
		\centering
		\includegraphics[width=1.0\linewidth]{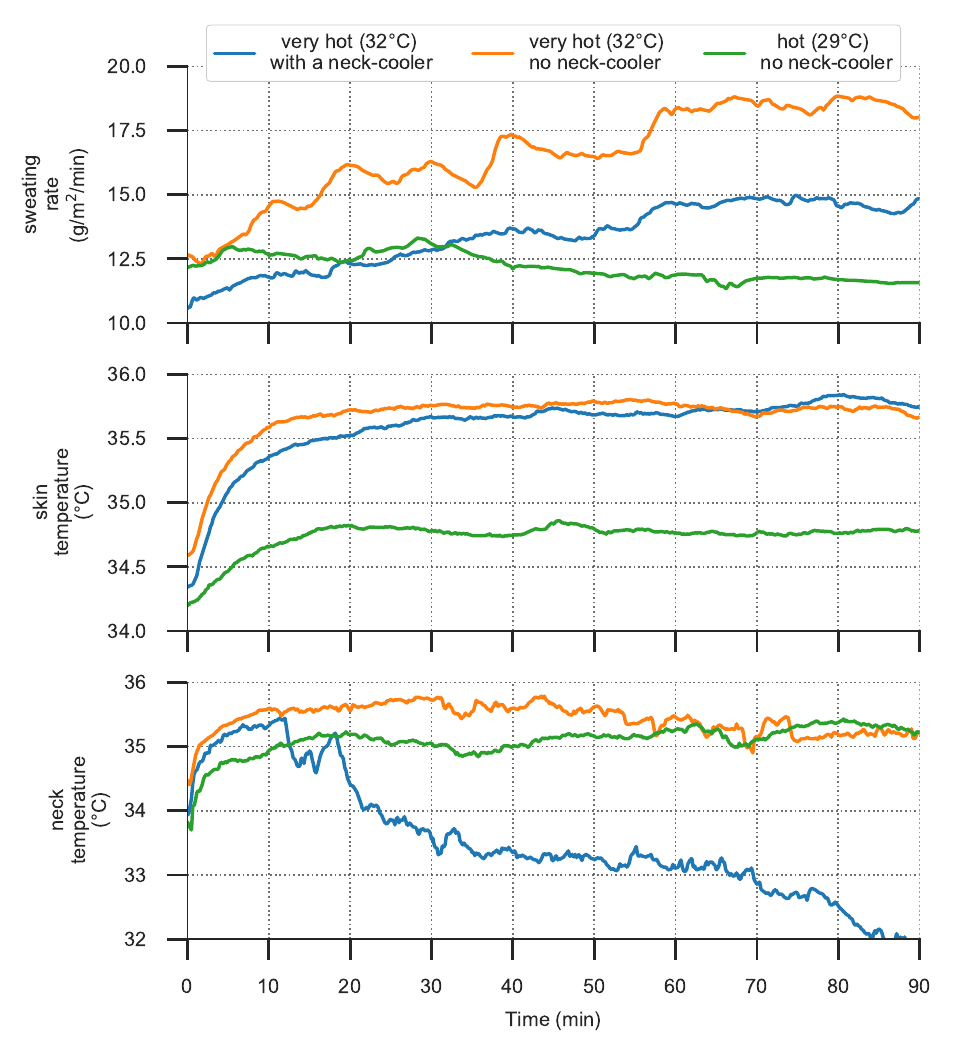}
		\caption{\textbf{Mean (11 subjects, 8350 samples) temperatures and sweating rate} \textemdash The recorded sensor data indicator that the neck-coolers reduced all subjects' skin temperature, and  their neck temperature $(\textit{P}<0.001)$. \\
			However, for the sweating rate, only 8 subjects' sweat rate is statistically significant $(\textit{P}<0.05)$.} 
		\label{fig:personaltemperature}
	\end{figure}
	\item 
	As for the subjects' HRV(\autoref{fig:hrv-summary-plot}), as we had previously shown in a different experiment \cite{Nkurikiyeyezu2017a}, the heart rate (HR) increased in hottest environments, the pNN50 was lowest in the hottest environment and the Very Low Frequency (VLF) band in the HRV power spectrum was highest in the coldest environment $(\textit{P}<0.001)$. In particular, when the subjects wore the neck-coolers, their pNN50 signal is consistently and conspicuously higher than their counterpart pNN50 signal when they did not wear any neck-coolers. According to our previous research \cite{Nkurikiyeyezu2017a}, this observation implies that, with the neck-coolers, subjects' heart beating patterns reflected that of a lower-temperature environment. Other HRV indices (especially the HR) seem to follow this pattern $(\textit{P}<0.001)$, albeit abstrusely and inconsistently.
	\begin{figure}[htbp]
		\centering
		\includegraphics[width=1.0\linewidth]{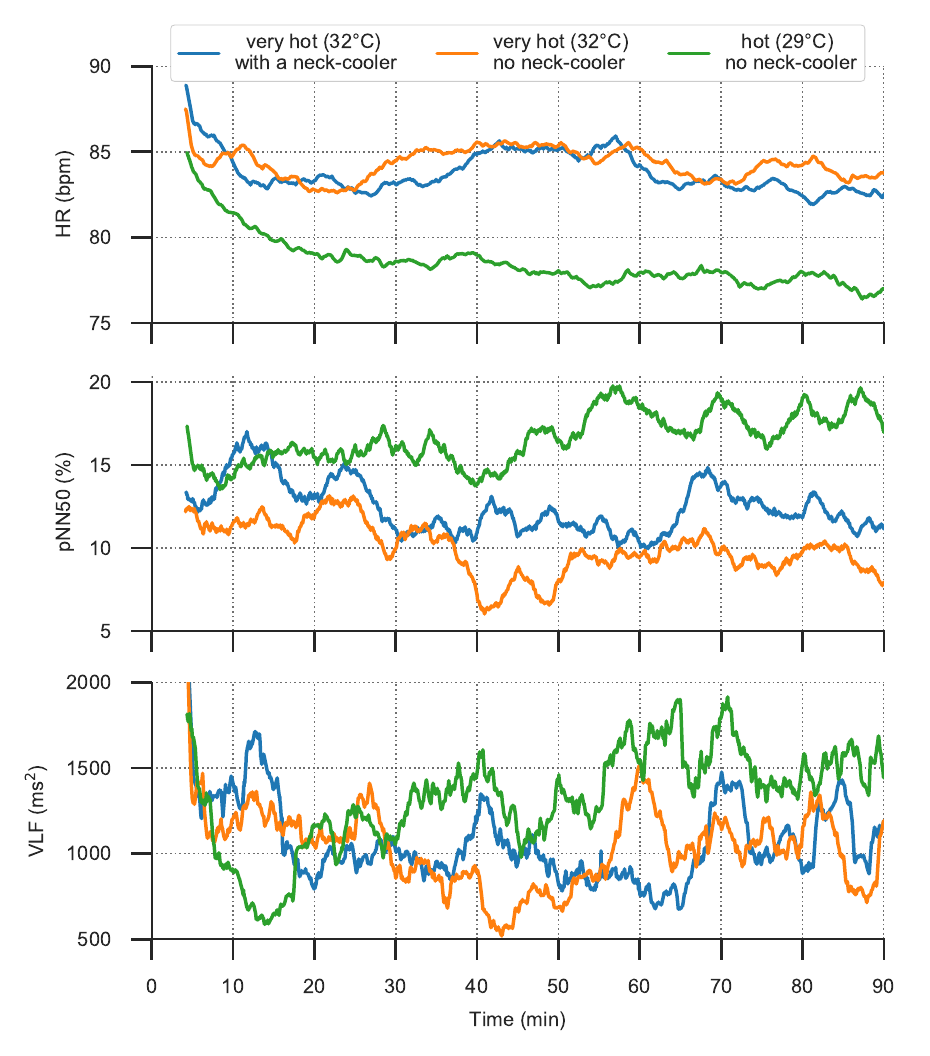}
		\caption{\textbf{Mean (11 subjects, 94094 samples) Heart Rate Variability (HRV)}\textemdash The HRV changes depending on the thermal environment. Strikingly, it seems that, when the subjects wore the neck-coolers, their heart beating patters (especially for the pNN50) were similar to the heartbeat pattern they would have if they were in slightly lower temperature environment.}
		\label{fig:hrv-summary-plot}
	\end{figure}
\end{itemize}
\subsection{Thermal comfort prediction}\label{sec-comfort-prediction}
\begin{figure}[tbph]
	\centering
	\includegraphics[width=1.0\linewidth]{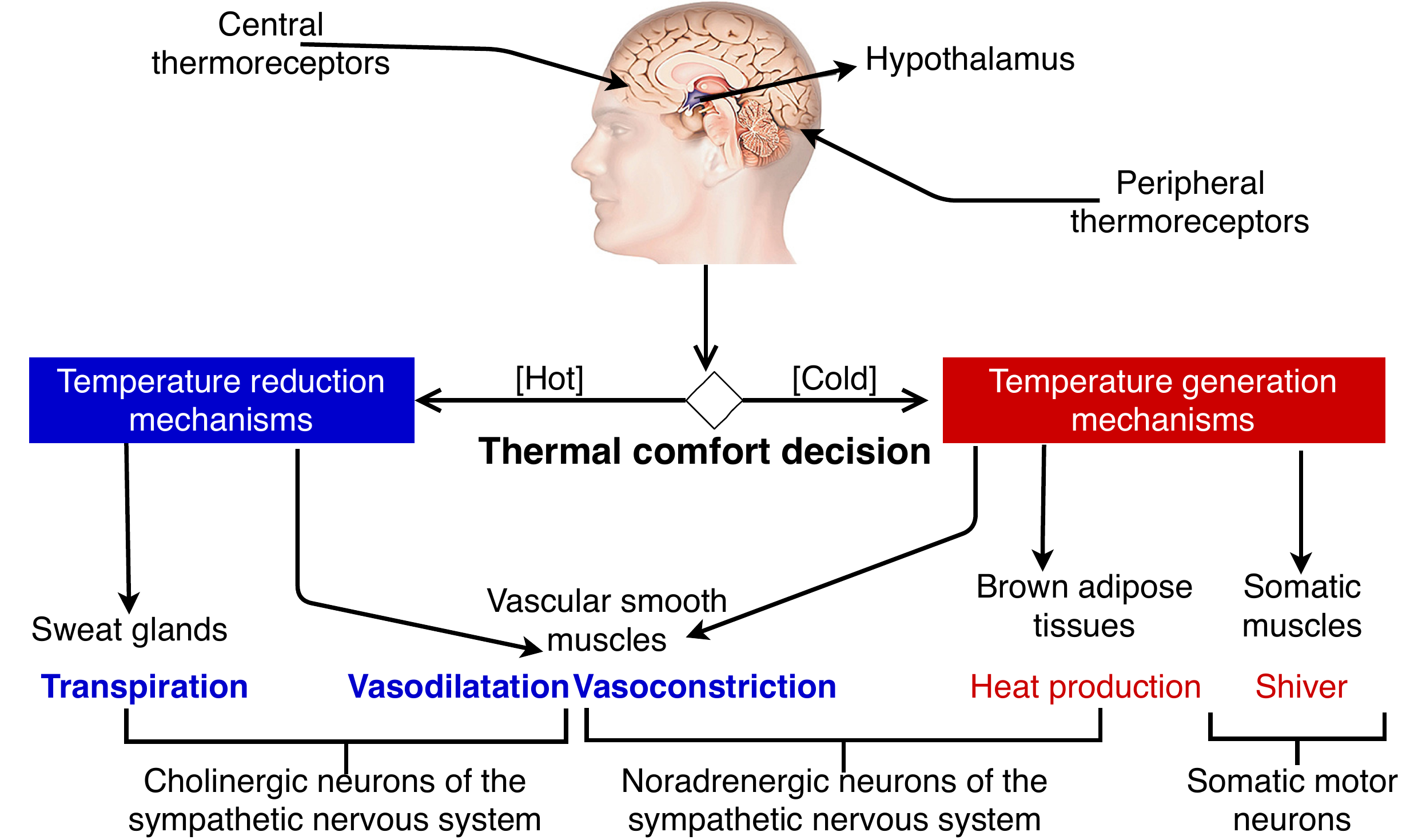}
	\caption{\textbf{A simplified illustration of thermoregulation} \newline The brain’s hypothalamus checks a person's core temperature and kickoff the thermogenesis or heat dissipating processes depending on whether the person feels hot or cold. }
	\label{fig:thermoregulation}
\end{figure}

The nervous system is a crucial player in people's thermoregulation\cite{Morrison2011}. For example, a reduction in a person's skin temperature leads to thermogenesis, shivering, and an increase in his neck muscle activities. On the contrary, when the skin temperature increases, there is vasodilation and transpiration to bolster heat loss  (\autoref{fig:thermoregulation}). From this observation, we had previously shown \cite{Nkurikiyeyezu2017a} that it is possible to apply machine learning algorithms on a person's fluctuation in his physiological signals and to predict his thermal comfort in real-time \cite{Nkurikiyeyezu2018b}.
\par
\begin{figure}[tbph]
	\centering
	\vspace{-4mm}
	\includegraphics[width=1.0\linewidth]{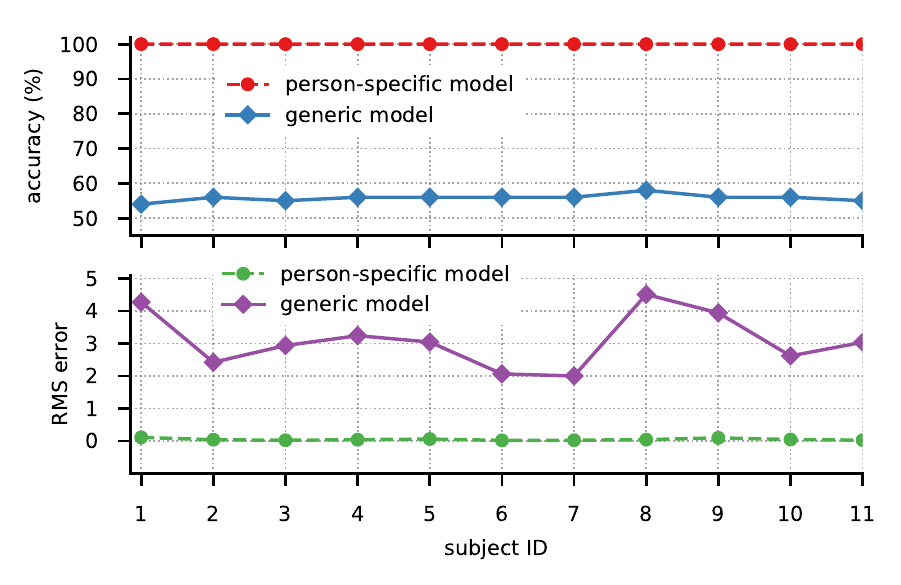}
	\caption{\textbf{Performance of the person-specific versus the generic model} \newline For all subjects (N=11 subjects), the person-specific models achieved a near-perfect prediction. However, due to the differences in how each subject physiologically responds to thermal comfort, the generic model performed crudely.}
	\label{fig:generic-vs-pers-spec-model}
	\vspace{-2mm}
\end{figure}

 
We applied regression and classification machine learning algorithms on the HRV dataset as described in \autoref{model-validation}. We found that person-specific models achieved a suspiciously too good to be true performance ($accuracy=99.98\pm 0.01$, $RMSE=0.04 \pm 0.01$)\footnote{ \textcolor{Gray}{see Table II and Table III in the \suppmaterial.}}. As shown in \autoref{fig:generic-vs-pers-spec-model},  this superb performance is, however, only half the story of the real performance of the models\footnote{ \textcolor{Gray}{see Table IV and Table V in the \suppmaterial.}}. Indeed, when we tested the same models on the unseen subjects (i.e., the subjects whose HRV samples were not part of the training set), the performance significantly decreased (\text{$accuracy=55.8\pm 0.98 \%, RMSE=3.50\pm 0.83, R^2 <0$}). 
\par 
As unlikely as it seems, the models' poor performance on the unseen subjects (i.e., the subjects whose data were not used for training the model) is arguably not a fruit of model overfitting because of four reasons. First, we used a large training dataset (364788 samples, 66 HRV features, size $\approx 430MB$). Although a large training set does not override overfitting, in most cases, a large train set is more diverse and reduces overfitting. Second, we tested numerous ensemble machine learning algorithms (e.g., \autoref{tab:model-parameters}) of various complexity, and they achieved an outstanding cross-validation performance, but they flopped when tested on the unseen subjects. Even a weak person-specific model that underfits the dataset performed better than the best person-independent model. As an illustration, a person-specific classification adaptive boosting model with a decision stump (i.e., a one-level decision tree) achieved a 56\% accuracy. Its regression counterpart had 2.27 and 0.12 RMSE and \textrm{$R^2$} respectively. Third, all our models are based on ensemble machine learning algorithms, which, while they can overfit in some cases, are designed to reduce the likelihood of overfitting. Finally, and most importantly, we evaluated the models using a cross-validation approach, and we obtained prediction with low standard deviations between the 10 folds.
\par 
The poor performance on the unseen subjects is, however, not completely unexpected. Thermal comfort is intrinsically an idiosyncratic psychological sensation that depends on factors that are unique to each person \cite{Wang2018b}. Thus, it is not possible for a model to generalize on new unseen subjects. We confirmed this phenomenon by investigating the influence of the individual differences to the performance of the models. We added a control prediction feature, the \textit{subject\_id}, to the datasets. The \textit{subject\_id} served as an indicator of the owner of each sample in the dataset.  We computed and compared the rank of all features in the dataset and found that the \textit{subject\_id} was always the most important feature. Similarly, we used a Recursive Feature Elimination (RFE) approach and found that the \textit{subject\_id} was always among the most important feature. 
\subsection{Thermal comfort model calibration}\label{model-calibration}
The poor performance of the generic model highlights the limitation of the system we had previously proposed. A practical thermal comfort prediction system that uses our approach needs to take into consideration the individual differences in how each person expresses thermal comfort. The system should, for example, use only person-specific models. This approach, however, is redundant, problematic, time-consuming, and very expensive to deploy in the real world because it would require to collect, label, and train extensive new data for every person in the building. Also, once the models are deployed, there is no guarantee they would work as expected because thermal comfort is affected by unpredictable factors \cite{Nicol2017, Wang2018b}. 
\par
\begin{figure}
	\centering
	\includegraphics[width=1.0\linewidth]{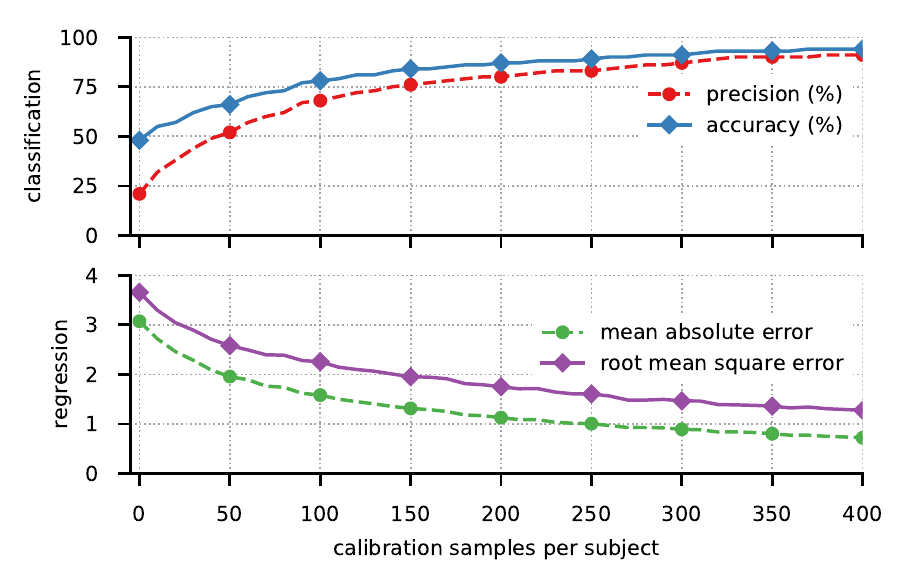}
	\caption{\textbf{Model calibration} \textemdash without the calibration samples, the models performed poorly. However, their performance steadily improved when they were calibrated with a few person-specific samples}
	\label{fig:model-calibration}
\end{figure}

 
In our previous studies \cite{Nkurikiyeyezu2019}, we proposed a model calibration technique which incorporates a few samples from unseen people into a generic model trained on the data of a large group of people. In a nutshell, the technique (\autoref{algo:model-calibration}) consists of adding a few person-specific calibration samples collected from previously unseen people into generic data collected from a large group of people. In this study, the success of the proposed method hinges in the fact that humans share a similar response to thermal discomfort \cite{Morrison2011}. However, every person possesses unique factors that supersede this generic response to thermal comfort \cite{Wang2018b}. From this observation, we hypothesis that adding a few person-specific calibration samples to the training data of the generic model would increase its performance because the new calibrated model would be able to capture the \say{uniqueness} of the new unseen people. When we applied this approach to the thermal comfort prediction, we observed that it improved the performance of the generic model considerably (\autoref{fig:model-calibration})\footnote{ \textcolor{Gray}{see Table VI and Table VII in the \suppmaterial.}}.
\begin{itemize}
	\item The classification model's performance steadily increased when the calibration samples were added. For example, the accuracy increased from 48\% to 82\% when we used 100 calibration samples and culminated in a 96\% accuracy when we used 400 calibration samples.
	\item 
	Similarly, for the regression model, the RMSE decreased from 3.65 to 1.09 when we used 400 calibration samples. At the same time, its  \textrm{$R^2$} coefficient improved from a futile \textrm{$R^2=-0.69$}  to a satisfactory \textrm{$R^2=+0.84$}
\end{itemize}
It is imperative to note that it only took a small fraction of the calibration samples to increase the performance of the generic models trained on a large dataset. In this study, we added 400 samples to dramatically increase the performance of the generic models that we had trained on approximately 270000 samples. For most people, it would only take 5 to 6 minutes to collect the required 400 HRV calibration samples.
\section{Thermal comfort provision}
\begin{figure}[htbp]
	\centering
	\includegraphics[width=1.0\linewidth]{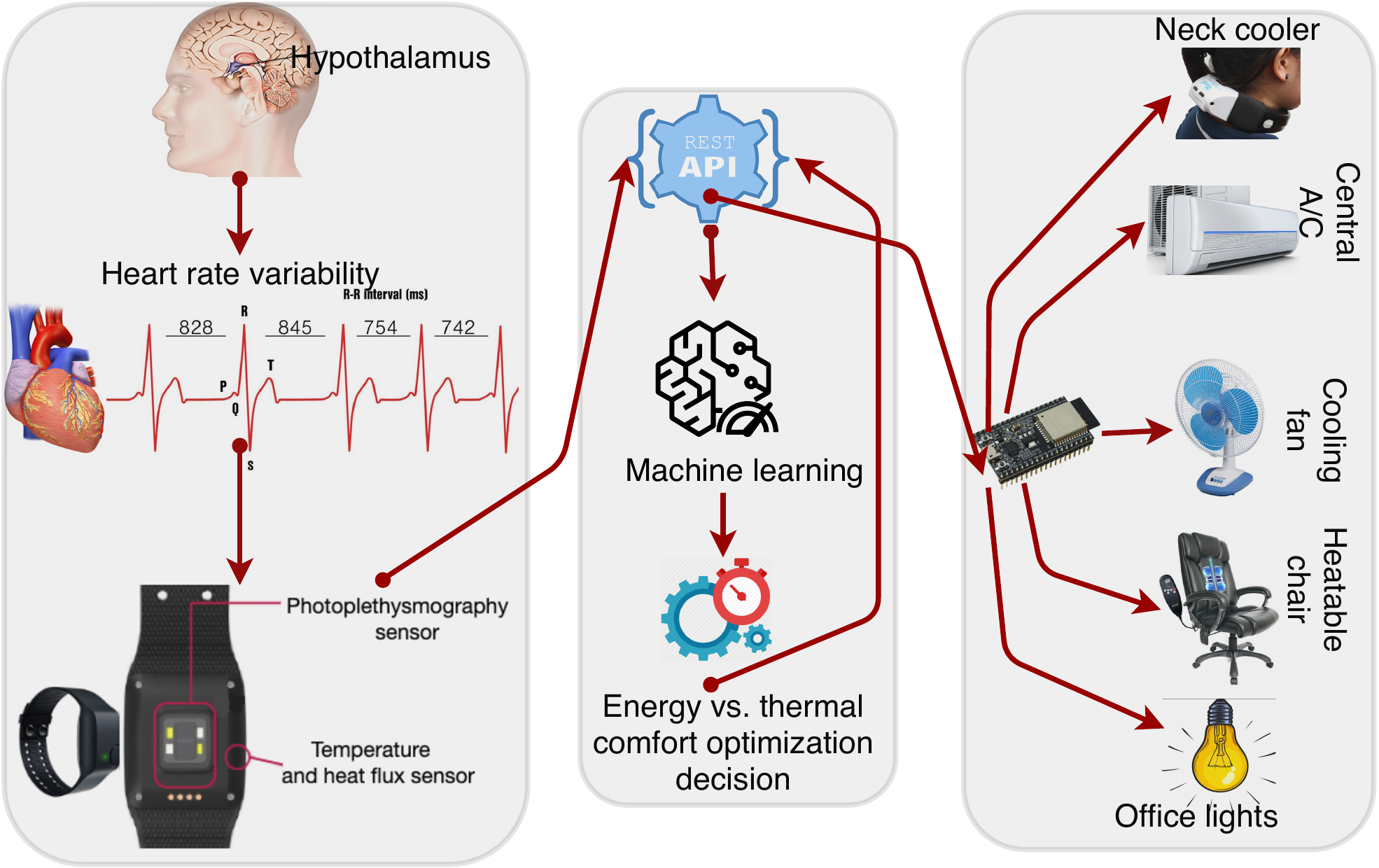}
	\caption{\textbf{Proposed comfort provision system}. \newline A person's photoplethysmogram (PPG) signal is recorded using a wristband device (e.g., an Empatica E4). The device periodically sends the PPG signal to a computing device which computes and pre-processes (e.g., data cleaning, re-balancing and dimension reduction) the HRV features and sent them to a remote server where they are used to predict the person's thermal comfort. The server then uses constrained optimization algorithms to activate suitable actuators (e.g., neck-cooler, a fan, and heated/cooled chair) to deliver an optimum and personalized thermal comfort using the least possible energy. In case the system miscalculates the person's thermal comfort, the user could adjust the temperature; thus, implicitly giving calibration samples that could be used to personalize the system to his liking further. This feedback is used to train a personalized stress prediction model, which is published and consumed as a RESTful API. When the model deteriorates,  it is automatically updated based on the periodic self-evaluations the system received from its users.}
	\label{fig:proposed-system}
\end{figure}

One of the pillars of responsive, intelligent buildings (IBs) is to respond to each occupant's need and to maximize his comfort and well-being with a minimal negative impact on the environment \cite{Ghaffarianhoseini2016}. Although there is a need for extensive studies to validate our findings, the results in this study corroborate with those in our previous studies and suggest that it could be possible to design an affect-aware thermal comfort provision system that self-adjust to meet the needs of every person in the building. Furthermore, as we discussed in our previous paper \cite{Nkurikiyeyezu2018a}, such a system would provide a higher quality thermal comfort and requires lower energy. An archetype of such a system is shown in \autoref{fig:proposed-system}. As already argued by other researchers\cite{Kim2018, Deng2016, Brager2015, Hoyt2014a, Pasut2015}, personalized comfort models do provide better thermal comfort and require lower energy. For example, Pasut and his co-authors \cite{Pasut2015} showed that a personalized heated/cooled chairs and a fan were able to provide satisfactory thermal comfort to 90\% of the users. Furthermore, the whole setup consumed 50\% less energy compared to a central air-conditioning system. Our proposed method improves their approach one step further because it directly estimates the thermal comfort from the person's physiological response to the thermal environment and adjusts the thermal environment accordingly. Furthermore, in case the system misjudge a person's thermal comfort, the person could adjust the temperature according to his liking. 
\par 
Although the proposed system has many advantages, its deployment in real-world settings poses numerous challenges. Notably, there is a need to collect the bio-signals unobtrusively, and its deployment would certainly need substantial upfront investment. Fortunately, there exist many enabling technologies that would ease some of these challenges. For example, it could be possible to use an Empatica E4 wristbands\footnote{\textcolor{Gray}{https://www.empatica.com/research/e4}} to collect the PPG signals. Furthermore, it might be economical to use off-the-shelf heatable ergonomic chairs, neck coolers, and fans instead of designing custom-made components. Finally, the design of the system might take advantages of the commercial machine learning cloud frameworks to simplify the deployment and maintenance of the machine learning models. For example, in our proposed system (\autoref{fig:proposed-system}), the designer could use the IBM's Watson Studio\footnote{\textcolor{Gray}{https://www.ibm.com/cloud/machine-learning}} to manage the thermal comfort models, including model calibration and deployment as a REST API. Additionally, even though the deployment and maintenance of the proposed system would require a significant upfront investment, the investment might pay off itself because of the expected energy-saving it would bring. Furthermore, the system could also double as a multipurpose system that uses the office occupants' physiological signals for preventive medicine and stress management. The advantages of these spillovers applications (preventive medicine and stress monitoring) are good enough alone to offset the cost of the initial investment because stress and sickness cost employers billions of dollars to compensate for their workers' sick leaves, lower productivity, job absenteeism, and high employee turnover. 
\section{Conclusion}
The technologies for thermal comfort provision in the buildings are ineffective in terms of the quality of the comfort they provide and in terms of the energy they require. Our research hinges on neuroscience and cardiology studies that showed that thermally dis-comfortable environments yield detectable changes in a person's physiological signals, e.g., the variability in a person's heart beat-to-beat intervals. We used these breakthrough to develop an intelligent thermal comfort provision system that delivers the thermal comfort depending on the fluctuations in a person's photoplethysmogram (PPG) signal. This system, however, used a generic thermal comfort prediction model and did not take into account the uniqueness in how each person expresses thermal comfort.
\par 
In this paper, we conducted experiments on 11 subjects doing office work in three thermal environments and compared how their thermal comfort would improve when they wore neck-coolers.  We found that the subjects expressed having a higher thermal comfort when they wore the neck-coolers.  Therefore, using neck-coolers in the summers might save energy by keeping the indoor temperature at relatively high temperatures without compromising the occupants' thermal comfort. 
\par
Moreover, because thermal comfort is a subjective psychophysiological feeling that is influenced by many unpredictable factors, generic thermal comfort prediction models cannot work well. Instead, person-specific models do. Regrettably, the deployment of a system that uses person-specific models would be costly and inflexible for deployment into real-world buildings. In this paper, we proposed a model calibration technique that incorporates a few calibration samples into the training data of a generic model to increase its performance. For example, when we added 400 calibration samples into a 270000 samples training set, its accuracy increased from 47.7\% to 96.1\%. We also discussed a possible implementation of a real-time thermal comfort provision system that uses our techniques and highlighted its advantages and limitations.
\scriptsize
\section*{Supplementary materials} \label{sec::supplement}
The supplemental material\footnote{\textcolor{Gray}{freely available at \datasetUrl}}contains:
\begin{itemize}
	\item Method\textemdash Detailed description of \cref{sec:method}
	\item Table I\textemdash List of the heart rate variability (HRV) features (\cref{sec:dataset})
	\item Table II and Table III\textemdash Performance of the person-specific models (\cref{sec-comfort-prediction})
	\item Table IV and Table V\textemdash Performance of the generic models (\cref{sec-comfort-prediction})
	\item Table VI and Table VII \textemdash Performance of the calibrated models (\cref{model-calibration})
	\item Table VIII and Table IX \textemdash Feature importance when the \textit{subject\_id} feature is added to the dataset (\cref{sec-comfort-prediction})
	\item Plots comparing the performance of the different machine learning models
	\item The computed HRV dataset (\cref{sec:dataset})
	\item The python source code to reproduce the key findings of this research
\end{itemize}
{
	\scriptsize
	\vspace{2cm}
	\linespread{0.9}
	\bibliographystyle{IEEEtran}
	\bibliography{clean-ACII2019.bib}
}
\end{document}